\documentclass{ws}
\newcommand{\citen}[1]{[\refcite{#1}]}
\newlength{\figurewidth}
\begin{document}

\title{\vspace*{-10mm}Analytic invariant charge: \\
       non-perturbative aspects}

\author{\vspace{-5mm}A.V.\ NESTERENKO
\footnote{\uppercase{W}ork partially supported by grant 00--15--96691
          of the \uppercase{RFBR}.}}

\address{Bogoliubov Laboratory of Theoretical Physics \\
Joint Institute for Nuclear Research, Dubna, 141980, Russia \\
E-mail: nesterav@thsun1.jinr.ru}

\maketitle
\vspace*{-10mm}
\section{Introduction}
     Theoretical analysis of hadron dynamics in main part relies upon
the renormalization group (RG) method. However, construction of exact
solutions to the RG equation is still far from being feasible.
Usually, in order to describe the strong interaction processes in
asymptotical ultraviolet (UV) region, one applies the RG method
together with perturbative calculations. The relevant approximate
solutions to the renormalization group equation are used for
quantitative analysis of high-energy experimental data. However, such
solutions contain unphysical peculiarities in the infrared (IR)
domain.

     An effective way to overcome such difficulties consists in
invoking into consideration the analyticity requirement, which
follows from the general principles of local Quantum Field Theory
(QFT). This idea became the basis of the so-called analytic approach
to QFT, which was first formulated in late 1950's \citen{RedBLS}.
Recently this approach has been extended to Quantum Chromodynamics
(QCD) \citen{ShSol,DV}.

\vspace{-2.5mm}
\section{New analytic invariant charge}
     The perturbative approximation for the $\beta$ function in
renormalization group equation for invariant charge $\alpha(\mu^2) =
g^2(\mu^2)/(4\pi)$
\begin{equation}
\frac{d\,\ln \bigl[g^2(\mu^2)\bigr]}{d\,\ln\mu^2} =
\beta\Bigl(g(\mu^2)\Bigr)
\end{equation}
leads to unphysical peculiarities of outcoming solutions (e.g.,
Landau pole). In the framework of developed model
\citen{PRD1,PRD2,NPQCD01} the analyticity requirement is imposed on
perturbative expansion of the RG $\beta$ function for restoring its
correct analytic properties.  At the one-loop level the corresponding
renormalization group equation can be solved explicitly:
\begin{equation} \alpha^{\mbox{\scriptsize (1)}}_{\mbox{\scriptsize
an}} (q^2) = \frac{4\pi}{\beta_0}\,\frac{z-1}{z\,\ln z}, \qquad z =
\frac{q^2}{\Lambda^2}.  \end{equation}

     At the higher loop levels only the integral representation for
the analytic invariant charge (AIC) was derived (see Refs.\
\citen{PRD2,MPLA2}). Figure~\ref{Fig:naichl} presents the analytic
running coupling $\widetilde{\alpha}_{\mbox{\scriptsize an}} (q^2) =
\alpha_{\mbox{\scriptsize an}} (q^2) \beta_{0}/(4\pi)$ at different
loop levels. The properties of the analytic invariant charge and the
relevant $\beta$ function are investigated in details in
Refs.~\citen{MPLA2,MPLA1}.

\vspace{5mm}

\noindent
\parbox[t]{\textwidth}{
\begin{tabular}{cc}
\parbox[t]{\figurewidth}{
\centerline{\epsfxsize=\figurewidth\epsfbox{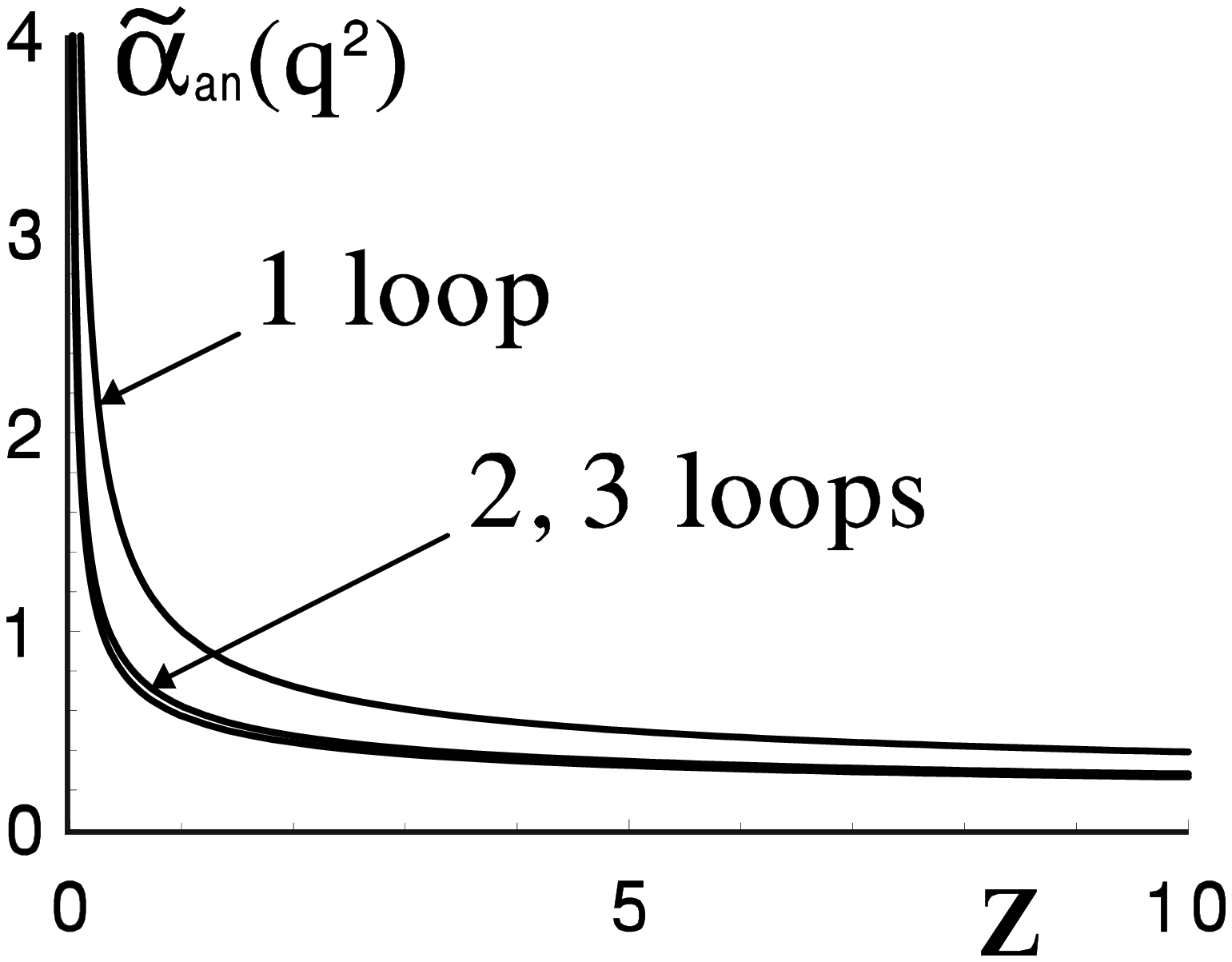}}
\refstepcounter{figure}
\label{Fig:naichl}
}
&
\parbox[t]{\figurewidth}{
\centerline{\epsfxsize=\figurewidth\epsfbox{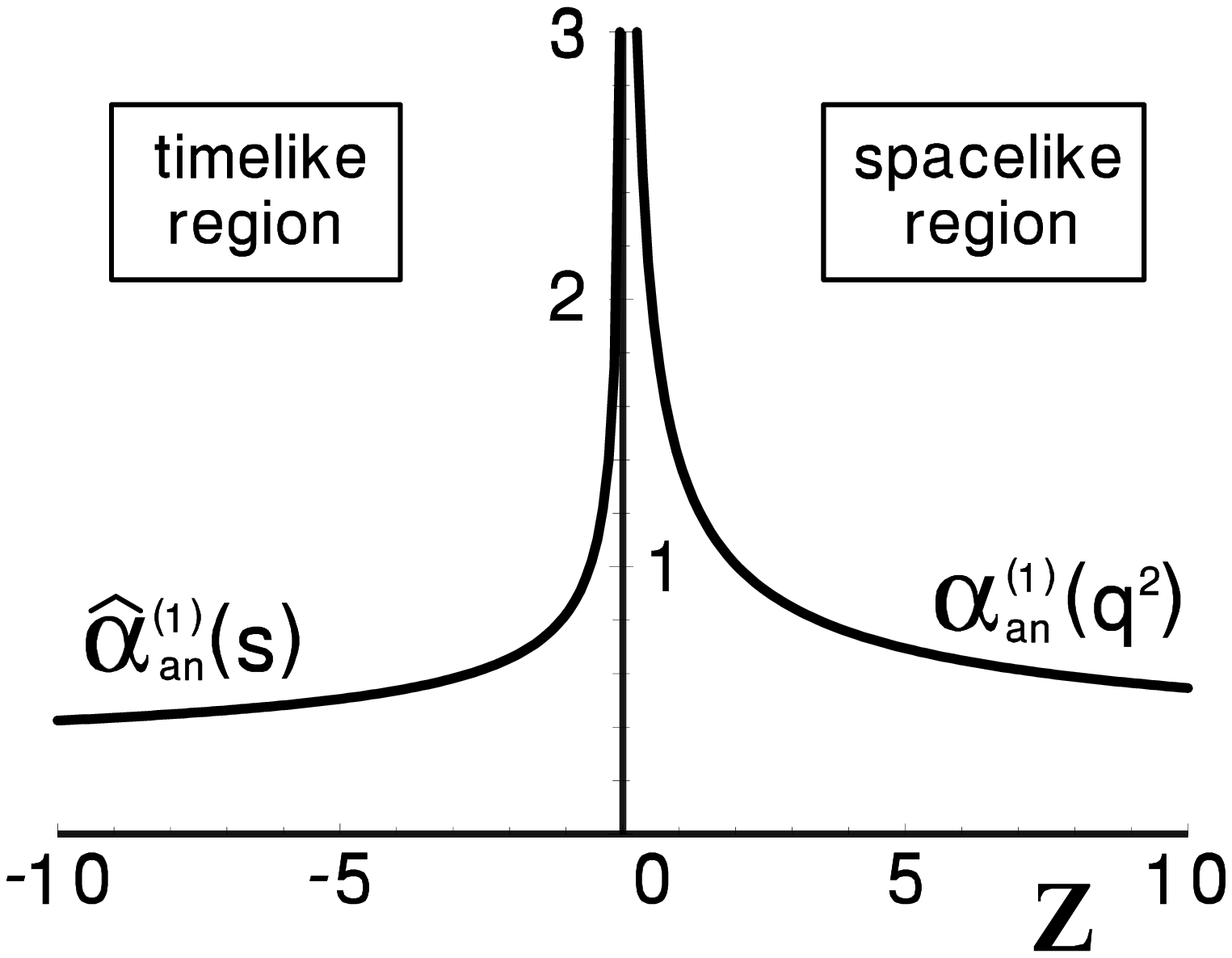}}
\refstepcounter{figure}
\label{Fig:naicst}
}
\end{tabular}
}
\noindent
\parbox[t]{\textwidth}{
\begin{tabular}{cc}
\parbox[t]{\figurewidth}{
{\scriptsize Figure 1. The analytic invariant charge at different
loop levels, $z=q^2/\Lambda^2$.

}}
&
\parbox[t]{\figurewidth}{
{\scriptsize Figure 2. The one-loop AIC in the spacelike and timelike
regions, $z=q^2/\Lambda^2$.

}}
\end{tabular}
}

\vspace{5mm}

     All stated above relate to the spacelike values of kinematic
variable ($q^2 > 0$). However, for the consistent description of some
hadron processes one has to employ the running coupling in the
timelike region \mbox{($s = -q^2 >0$)}. Recently the continuation of
the AIC to the timelike region has been performed \citen{PRD2}. In
particular, obtained result confirms the hypothesis due to Schwinger
concerning connection between the $\beta$ function and relevant
spectral density (see articles \citen{PRD2,MiltSol} and references
therein for the details). The plots of the one-loop AIC in the
spacelike and timelike regions are shown in Figure~\ref{Fig:naicst}.
In the ultraviolet limit these functions have identical behavior
determined by the asymptotic freedom. But there is asymmetry between
them in the intermediate energies region. The relative difference is
about several percents at scale of the $Z$ boson mass, and increases
when approaching the infrared domain.  Apparently, this circumstance
must be taken into account when one handles with experimental data.

\vspace{-4mm}
\section{Phenomenological applications}
     For verification of consistency of the model developed it is
worth  turning to its applications. Since we are working within a
non-perturbative approach, the study of the non-perturbative
phenomena is of a crucial importance.

     It has been shown \citen{PRD1} that the quark-antiquark
potential constructed by making use of the analytic invariant charge
is confining at large distances. The derived potential is in a quite
good agreement with lattice simulation data (see Refs.\
\citen{PRD1,NPQCD01} for the details).

     As known, some key non-perturbative aspects of strong
interaction are described by instantons. The distribution of
large-size instantons is directly related to the infrared behavior of
the invariant charge. The lattice simulation of this quantity
revealed severe suppression of the large-size instantons, that is
compatible neither with the perturbative results, nor with the
freezing of the QCD running coupling at large
distances~\citen{UKQCD}. Recently the conformal inversion symmetry of
this distribution was observed, which ultimately led to {\it
``{\hspace{0.7pt}}rediscovery''} of the analytic invariant charge
\citen{Schrempp}.

     The developed approach has also been applied to description of
gluon condensate, inclusive $\tau$ lepton decay, and
electron-positron annihilation to hadrons~\citen{NPQCD01}. The
consistency of estimated values of the scale parameter
(\mbox{$\Lambda_{\mbox{\tiny QCD}}\simeq 550\,$MeV}, one-loop level,
three active flavors) testifies that the analytic invariant charge
substantially incorporates, in a consistent way, both perturbative
and non-perturbative aspects of Quantum Chromodynamics.

\vspace{-2.5mm}
\section{Conclusion}
     The developed model for the QCD analytic invariant charge
possesses a number of profitable features. Namely, it has no
unphysical peculiarities at any loop level; it contains no free
parameters; it incorporates UV asymptotic freedom with IR
enhancement; it has universal behavior both in UV and IR regions at
any loop level; it possesses a good higher loop and scheme stability.
There is considerable difference between the respective values of AIC
in the low-energy domain of spacelike and timelike regions. The
developed model enables one to describe various strong interaction
processes both of perturbative and intrinsically non-perturbative
nature.

\end{document}